\documentclass[a4paper,11pt]{article}
\usepackage{pos}

\usepackage{amsmath}
\usepackage{siunitx}
\usepackage{todonotes}
\usepackage{subcaption}
\usepackage{multirow}
\usepackage{physics}
\usepackage{cleveref}

\newcommand{\eff}{\text{eff}}
\newcommand{\zth}{^{(0)}}
\newcommand{\fst}{^{(1)}}
\newcommand{\qcdiso}{$\text{QCD}_{\text{iso}}$}
\newcommand{\cratio}{\frac{c_*\zth}{c\zth}}
\newcommand{\expansioncoeff}{\cratio e^{-\Delta M\zth t}}

\title{Setting the Scale Using Baryon Masses with Isospin-Breaking Corrections}

\author*[a]{Alexander M. Segner}
\author[b]{Andrew D. Hanlon}
\author[c]{Andreas Risch}
\author[a,d,e]{Hartmut Wittig}

\affiliation[a]{Institut für Kernphysik, Johannes Gutenberg-Universität, \\
    Mainz, Johann-Joachim-Becher-Weg 45, 55128 Mainz, Germany}

\affiliation[b]{Physics Department, Brookhaven National Laboratory, \\
Upton, New York 11973, USA}

\affiliation[c]{John von Neumann-Institut für Computing NIC, Deutsches Elektronen-Synchrotron DESY,\\
    Platanenallee 6, 15738 Zeuthen, Germany}

\affiliation[d]{Helmholtz Institut Mainz, \\
    Staudingerweg 18, 55128 Mainz, Germany}

\affiliation[e]{Helmholtzzentrum für Schwerionenforschung, \\
    64291 Darmstadt, Germany}

\emailAdd{alsegner@uni-mainz.de}

\abstract{ We present the status of an ongoing project aimed at the
  inclusion of isospin-breaking corrections arising from the
  non-degeneracy of the light quark masses and electromagnetic
  interactions in calculations of the low-lying octet and decuplet
  baryon masses. Our ultimate goal is to perform a precision
  determination of the lattice scale including isospin-breaking
  effects. We apply the perturbative formalism to isospin-symmetric
  $N_f=2+1$ CLS ensembles. We show results on baryon masses up to
  leading order in isospin-breaking corrections on two ensembles with
  $m_\pi\approx\SI{290}{MeV}$ and $m_\pi\approx\SI{215}{MeV}$ at a
  lattice spacing of $a\approx\SI{0.076}{fm}$.

\flushright{MITP-22-101, DESY-22-198}}

\FullConference{%
The 39th International Symposium on Lattice Field Theory,\\
8th-13th August, 2022,\\
Rheinische Friedrich-Wilhelms-Universität Bonn, Bonn, Germany
}

\begin{document}

\maketitle

\section{Introduction}\label{sec:introduction}


We are in an era of precision lattice QCD, where contributions from QED and
strong isospin-breaking can no longer be ignored in calculations of many
phenomenologically relevant observables.
An example where these contributions are of significant importance is in the
lattice determination of the anomalous magnetic moment of the muon $(g-2)_\mu$.
One of the largest uncertainties in any lattice determination of $(g-2)_\mu$
comes from the scale setting \cite{Meyer:2018til}, and it is of great
importance that the physical scale is determined with sufficient accuracy (i.e.
below $1\%$), incorporating QED and isospin-breaking effects.

As lattice QCD ensembles are simpler to generate in the isosymmetric theory
\qcdiso\ than in full QCD+QED, many existing
ensembles are isospin-symmetric.
One method to incorporate isospin-breaking effects is based on their
perturbative treatment on isosymmetric ensembles
\cite{deDivitiis:2011eh,deDivitiis:2013xla}.
The goal of our project is to employ this formalism to perform a precision
calculation of the lattice scale for the $N_f=2+1$ CLS ensembles.

In refs. \cite{Bruno:2016plf,Strassberger:2021tsu}, the lattice scale for the
CLS ensembles was determined using a combination of pion and kaon decay
constants.
While the final result has a total error at the level of 1\%, the incorporation
of isospin-breaking corrections turns out to be conceptually quite difficult
\cite{Carrasco:2015xwa}.
In this project, we investigate the prospects of precision scale setting using
the masses of the lowest-lying baryon octet and decuplet, for which
isospin-breaking effects are simpler to incorporate.
In this contribution we present first preliminary results for our most
promising candidates for the reference scale on two CLS ensembles, N451 and
D450, with anti-periodic temporal boundary conditions.

We begin by giving an overview of our simulation setup and the isospin-breaking
expansion of baryonic two-point functions in \cref{sec:setup}.
We then continue with an introduction of our fit strategy in
\cref{sec:spectroscopy} and a discussion of the practical aspects and first
insights of our analysis in \cref{sec:analysis}.
Finally, a summary of the most important points and our plans for the next
steps in this project are given in \cref{sec:conclusion}.

\section{Simulation Setup and Expansion in Isospin-Breaking Parameters}\label{sec:setup}

Our analysis makes use of the CLS $N_f=2+1$ ensembles \cite{Bruno:2014jqa} with
$\mathcal O(a)$-improved Wilson fermions \cite{Luscher:1996sc} and a tree-level
Lüscher-Weisz gauge action \cite{Bulava:2013cta}.
Our choice of quark sources are SU(3)-covariantly Wuppertal-smeared
\cite{Gusken:1989qx} point sources.
We apply APE smearing \cite{APE:1987ehd} to the gauge links. 
The smearing parameters were chosen such that they minimize the nucleon
effective mass at an early time on the H105 ensemble and that the smearing
radius as defined in \cite{UKQCD:1993gym} is approximately \SI{0.5}{fm}.

The baryonic operators used for this project follow a construction introduced
by the Lattice Hadron Physics Collaboration \cite{Basak:2005ir}.
This construction yields a number of operators for each octet and decuplet
baryon state, resulting in correlator matrices for each baryon.
For a more detailed description of our implementation of these operators, we
refer to \cite{Segner:2021yqo}.

For improved precision, we use all-mode-averaging
\cite{Blum:2012my,Blum:2012uh,Shintani:2014vja} with 32 sloppy sources per
gauge-con\-fig\-u\-ra\-tion and one source for bias-correction on the ensembles
discussed in \cref{sec:analysis}.

In order to determine baryon masses, we consider baryonic two-point functions
of the form $C(t)=\ev{\mathcal B(t)\overline{\mathcal B}(0)}$ computed from
zero-momentum projected baryonic creation and annihilation operators
$\overline{\mathcal{B}}=\overline{B}\overline{\Psi}\overline{\Psi}\overline{\Psi}$
and $\mathcal{B}=B\Psi\Psi\Psi$, where $\overline B$ and $B$ encode the spinor,
flavor, and color structure of the operators as well as their time-dependence.
As we use isospin-symmetric gauge ensembles, we include isospin-breaking
effects in terms of an expansion of full QCD+QED two-point functions about
their counterparts in \qcdiso.
This procedure was first introduced by the RM123 collaboration
\cite{deDivitiis:2011eh,deDivitiis:2013xla} as an expansion in terms of the
electromagnetic coupling $e$, the deviations $\Delta m_u$ and $\Delta m_d$ of
the light quark masses from their isospin-symmetric counterpart $m_{ud}$ as
well as a shift in the inverse strong gauge coupling $\Delta\beta$ and the
strange quark mass $\Delta m_s$.
In the following, we will collect these coefficients into a vector
$\Delta\varepsilon$ which is to be understood as the difference of a vector
$\varepsilon=(\beta,e^2,m_u,m_d,m_s)$ parameterizing QCD+QED and the
corresponding vector $\varepsilon\zth=(\beta,0,m_{ud},m_{ud},m_s^{(0)})$
parameterizing \qcdiso\ and whose entries we denote by $\Delta\varepsilon_i$.
Note that we explicitly assume $\Delta\beta=0$ throughout.

For baryonic correlation functions as defined above, this expansion takes the
form
\begin{align*}
    C^\varepsilon(t)=&C^{\varepsilon\zth}(t)
    +\sum_f\Delta m_f\pdv{C^\varepsilon(t)}{m_f}\eval_{\varepsilon=\varepsilon\zth}
    +e^2\pdv{C^\varepsilon(t)}{e^2}\eval_{\varepsilon=\varepsilon\zth},
\end{align*}
where the superscripts $\varepsilon$ and $\varepsilon\zth$ indicate that the
correlation functions are computed in QCD+QED or \qcdiso\ respectively.
In a diagramatic representation, the above expansion can be written as
\begin{align*}
    \ev{\vcenter{\hbox{\includegraphics[width=2.8cm]{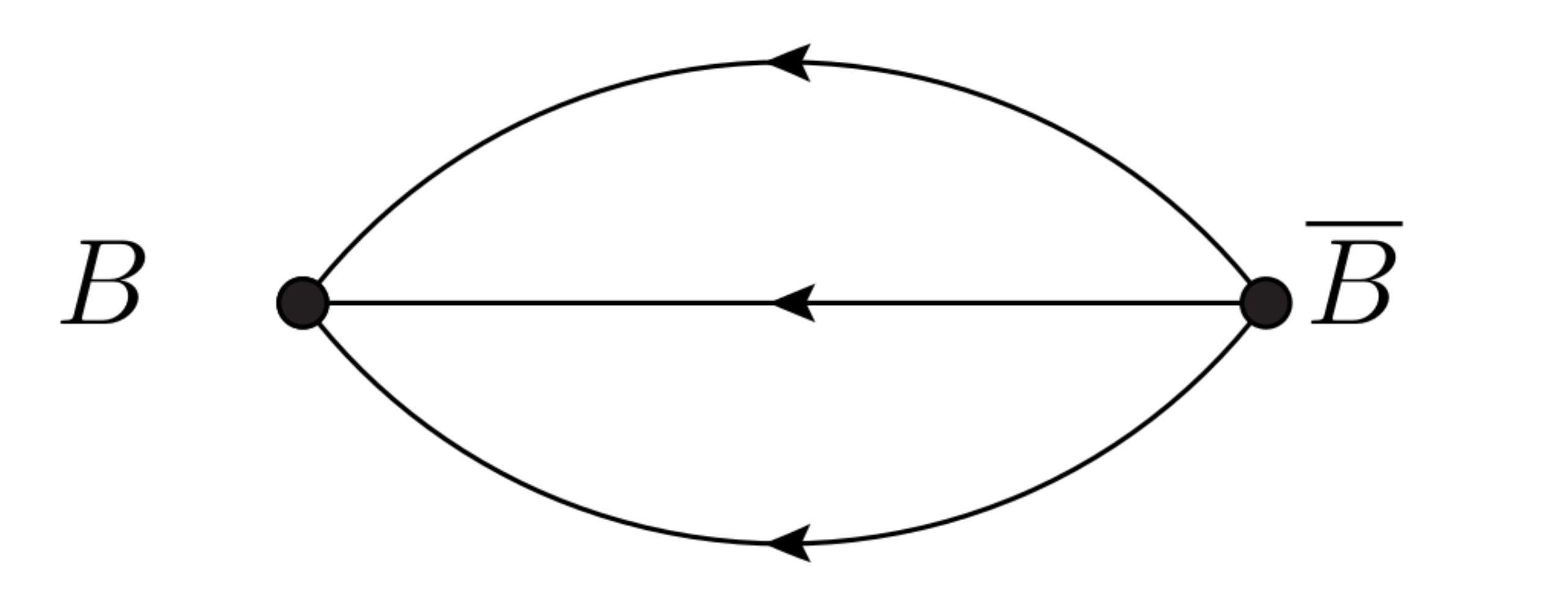}}}}^{\varepsilon}=&
        \ev{\vcenter{\hbox{\includegraphics[width=2.8cm]{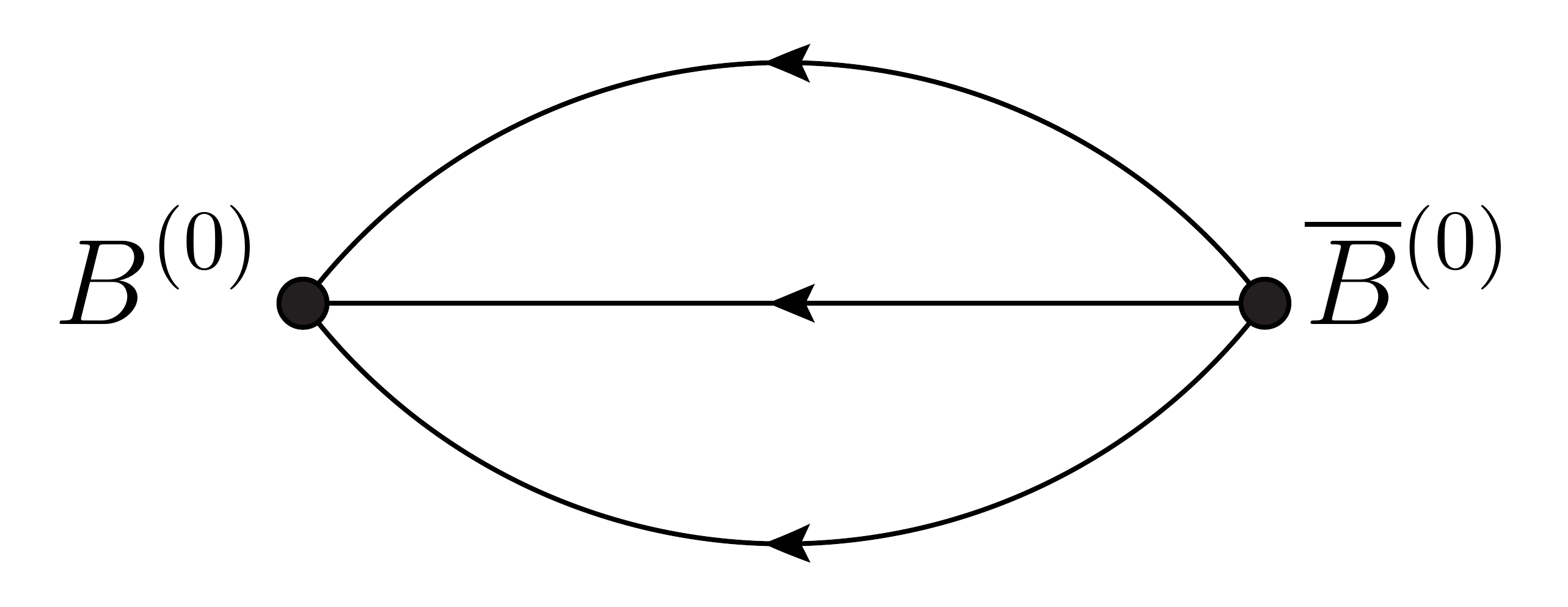}}}}^{\varepsilon^{(0)}}+
        \sum_f\Delta m_f\ev{\vcenter{\hbox{\includegraphics[width=2.8cm]{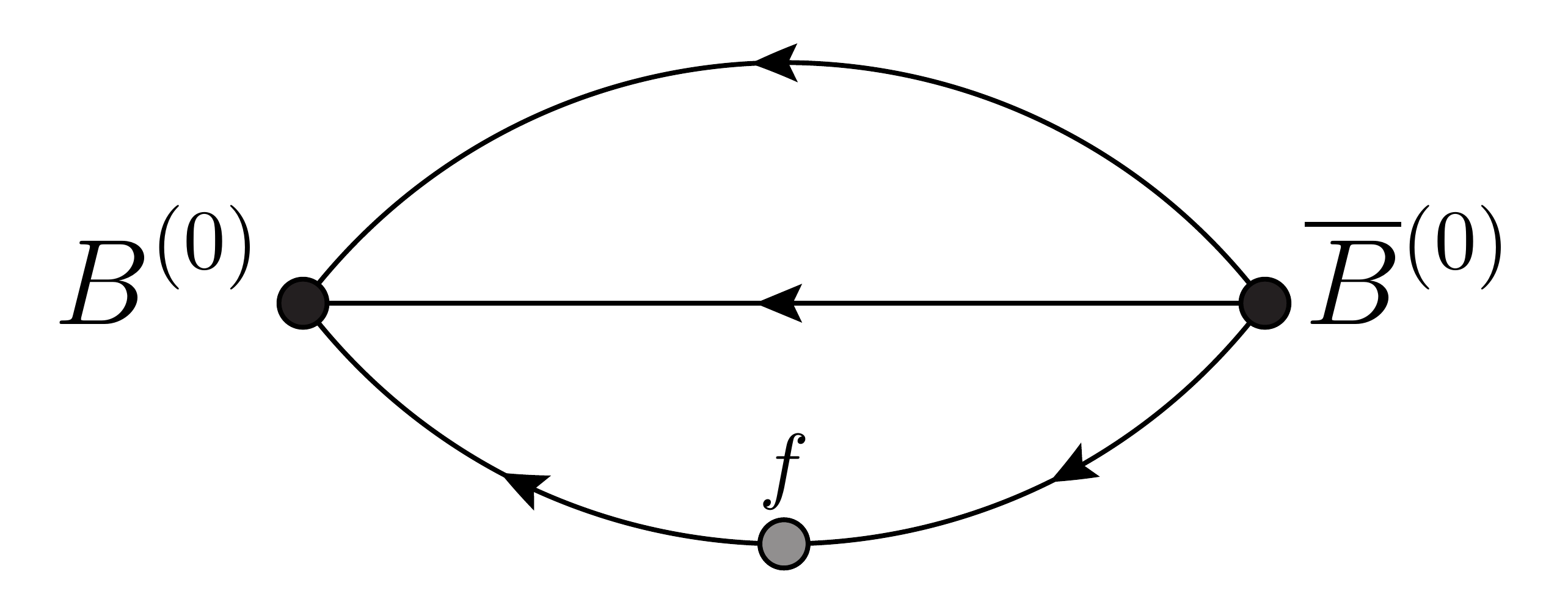}}}}^{\varepsilon^{(0)}}
        \\
        &+ e^2\ev{\vcenter{\hbox{\includegraphics[width=2.8cm]{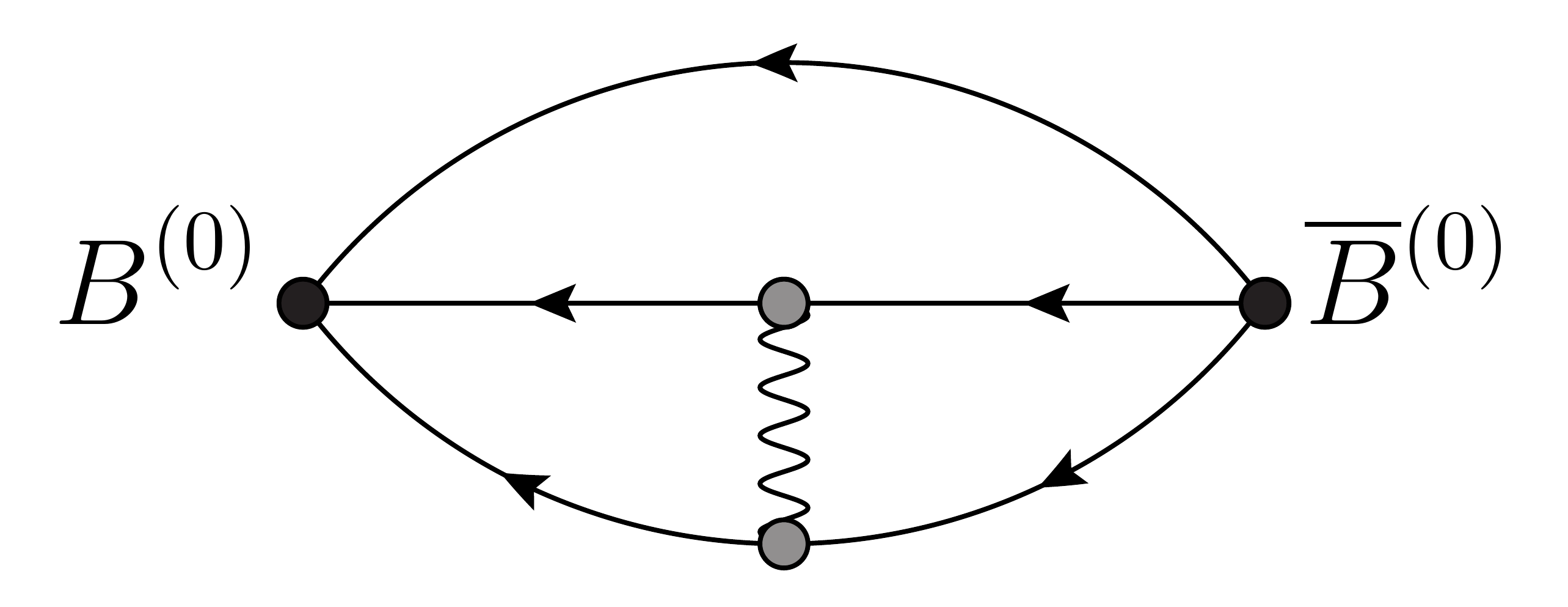}}} +
        \vcenter{\hbox{\includegraphics[width=2.8cm]{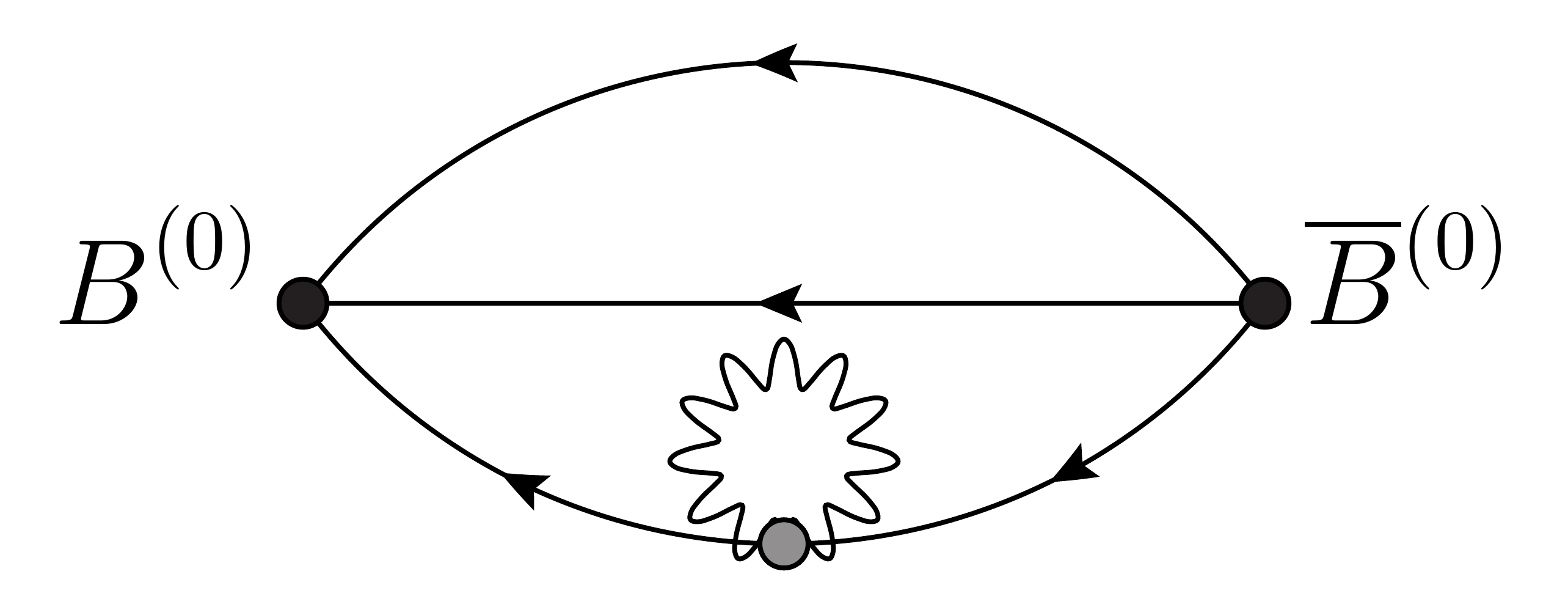}}} +
        \vcenter{\hbox{\includegraphics[width=2.8cm]{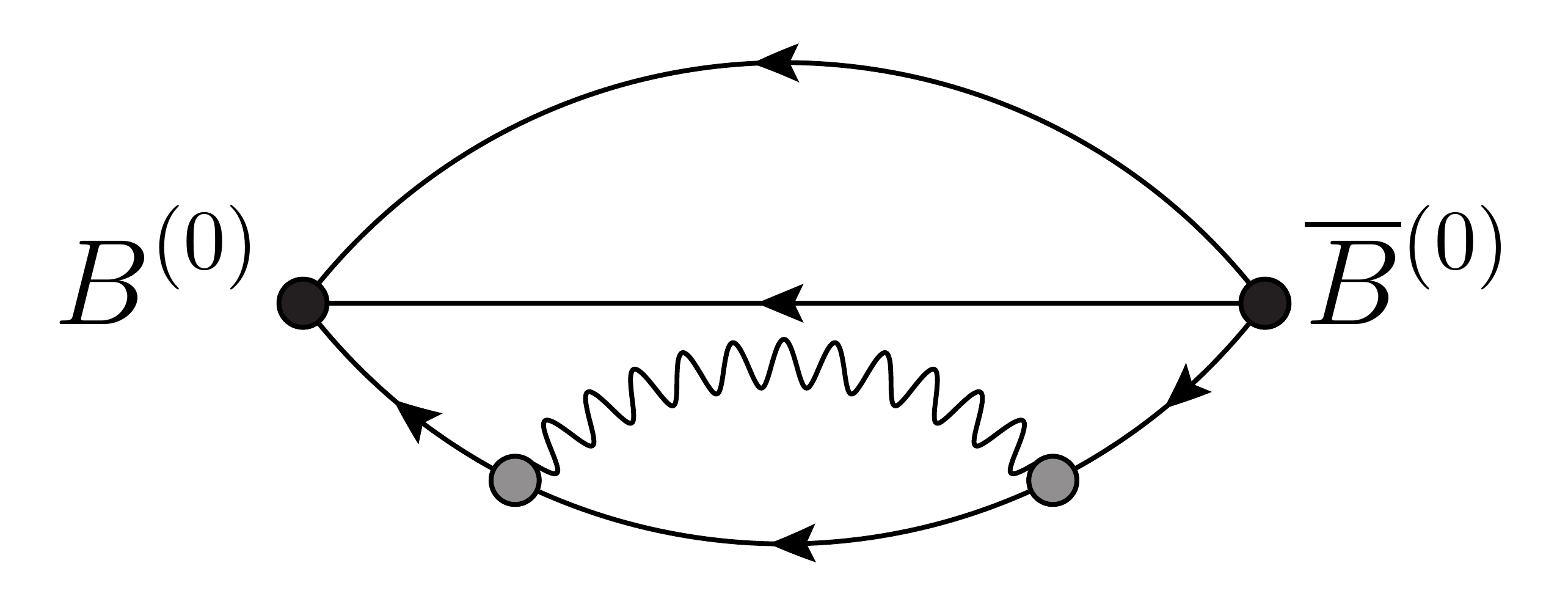}}}}^{\varepsilon^{(0)}}
        \\
        &+\cdots
\end{align*}
This expansion shows only those contributions which we consider in this study,
i.e. only the quark-connected contributions.  The complete first order in
isospin-breaking parameters furthermore contains corrections in the sea-quark
sector \cite{deDivitiis:2013xla}, which we do not include in our calculations.
We do, however, compute diagrams containing a single photon vertex with an
open-ended photon line which we can use later if we decide to include photon
exchanges between valence and sea quarks.
The correction terms are implemented using the method of sequential
propagators.

\section{Baryon Spectroscopy at Leading Order in Isospin-Breaking Corrections}\label{sec:spectroscopy}

The baryon masses are extracted from the two-point functions using the
asymptotic description $C(t)=ce^{-mt}$ which holds for large time-separations
$t$ between source and sink.
Using a perturbative approach to incorporate isospin-breaking effects, the
two-point functions must be expanded to the desired order to extract the
corrections to the baryon masses.
Expanding a general correlation function and its asymptotic form in terms of
the isospin-breaking parameters $\Delta\varepsilon$ up to leading order via
$X=X\zth+\sum_i\Delta\varepsilon_iX_i\fst+\mathcal O(\Delta\varepsilon^2)$
where $X\in\{c,m,C\}$ and
$X_i\fst:=\pdv{X}{\varepsilon_i}\eval_{\varepsilon=\varepsilon\zth}$, and
matching the zeroth and first order, one finds
\begin{align*}
    C\zth(t)=&c\zth e^{-m\zth t}, \\
    C_i\fst(t)=&\qty(c_i\fst-c\zth m_i\fst t)e^{-m\zth t}.
\end{align*}
From these expressions, the masses at zeroth and first order in
$\Delta\varepsilon$ can be extracted as
\begin{align}
\begin{aligned}\label{eq:meff_definition}
    m\zth=&-\dv t\log(C\zth(t)), \\
    m_i\fst=&-\dv t\frac{C_i\fst(t)}{C\zth(t)}.
\end{aligned}
\end{align}
For our analysis we use the following discretizations for these expressions as
definitions for the effective masses of the baryons:
\begin{align}
\begin{aligned} \label{eq:meff}
    (am_\eff)\zth(t):=&\log(\frac{C\zth(t)}{C\zth(t+a)}), \\
    (am_\eff)_i\fst(t):=&\frac{C_i\fst(t)}{C\zth(t)}-\frac{C_i\fst(t+a)}{C\zth(t+a)}.
\end{aligned}
\end{align}
The expressions in \cref{eq:meff} approach a plateau as $t\to\infty$.
In practice, however, excited states affect the effective mass at early times.
This would in principle not be much of a problem if there were little noise in
the region where the excited state effects become negligible.
However, since baryonic two-point functions suffer from signal-to-noise ratios
growing exponentially with $t$ \cite{Lepage:1989hd,Wagman:2016bam}, it is often
difficult to determine the position of this region.
Thus, to mitigate these deviations from a simple plateau, we include one
excited state in addition to the ground state, in which case the correlation
function takes the form
\begin{align}
    C_{2s}(t)=ce^{-mt}+c_*e^{-m_*t}\label{eq:2state_ansatz},
\end{align}
where $m$ is the ground state mass and $m_*$ is the mass of the first excitation.
Inserting \cref{eq:2state_ansatz} into \cref{eq:meff_definition} one finds
\begin{align}\label{eq:2state_fit0_exact}
    (am)_\eff\zth(t)=\frac{am\zth+\cratio am_*\zth e^{-\Delta M\zth t}}{1+\cratio e^{-\Delta M\zth t}}
\end{align}
where $\Delta M\zth=m_*\zth-m\zth$.
Following Del Debbio et al. \cite{DelDebbio:2006cn} we use a fit function
obtained from \cref{eq:2state_fit0_exact} by expansion in terms of $\expansioncoeff$:
\begin{align}
    (am)\zth_{\text{eff}}(t)=am\zth+\cratio a\Delta M\zth e^{-\Delta M\zth t}
                            =:am\zth+\gamma e^{-\Delta M\zth t}. \label{eq:2state_fit0}
\end{align}
Performing the expansion in isospin-breaking parameters for the
two-state ansatz (\cref{eq:2state_ansatz}) and plugging the result into the
first order formula in \cref{eq:meff_definition}, one finds a similar fit
function by expansion in terms of the same quantity as to obtain
\cref{eq:2state_fit0}:
\begin{align}\label{eq:2state_fit1}
    (am)_\eff\fst=am\fst+(\alpha-\beta t)e^{-\Delta M\zth t} ,
\end{align}
where
\begin{align*}
    \alpha=&\gamma\qty(\frac{\Delta M\fst}{\Delta M\zth}-\frac{c\fst}{c\zth}+\frac{c_*\fst}{c_*\zth}), \\
    \beta=&\gamma a\Delta M\fst, \\
    \Delta M^{(k)}=&m_*^{(k)}-m^{(k)}.
\end{align*}
Notice that the exponent in \cref{eq:2state_fit1} is the same as that in
\cref{eq:2state_fit0}.
This reduces the number of fit parameters in the first order fits as one can
simply reuse the fit results for $\Delta M^{(0)}$ from the zeroth order fits.

For the fits designed to determine the isospin-symmetric contribution, an
expansion in terms of $\expansioncoeff$ is not strictly necessary as the exact
expression (\cref{eq:2state_fit0_exact}) is generally not more difficult to fit
than the approximation and doesn't introduce any further independent fit
parameters.
For the isospin breaking corrections, however, the exact expression has one
more independent parameter and the mass correction $m\fst$ cannot be
extracted in a straightforward way compared to the expansion in
\cref{eq:2state_fit1}.

\section{First Results}\label{sec:analysis}


For this project, we have processed two ensembles (N451 at
$m_\pi\approx\SI{215}{MeV}$ and D450 at $m_\pi\approx\SI{290}{MeV}$) with
a lattice spacing of $a\approx\SI{0.076}{fm}$ (determined in the
isospin-symmetric theory \cite{Bruno:2016plf} using combinations of the
decay constants $f_\pi$, $f_K$, and the scale quantity $t_0$ derived from
the Wilson flow \cite{Luscher:2010iy}).
Both of these ensembles have antiperiodic temporal boundary conditions with
volumes of $128\times48^3$ for N451 and $128\times64^3$ for D450.

The analysis on these ensembles is based on the fit models described in
\cref{sec:spectroscopy} with the two-point functions calculated as described in
\cref{sec:setup}.
Our reasoning to study excited state effects in our data arise from the problem
of identifying a plateau in noisy data, which requires that excited-state
effects are sufficiently suppressed relative to the statistical fluctuations.
In addition to giving a clearer picture as to where a plateau might start, a
two-state fit often gives a more reliable estimate of the mass of the ground
state than a single-state fit.
Especially in fits to isospin-breaking corrections, where the extent of the
plateau is not always clearly recognizable due to quickly growing uncertainties
towards larger $t$, the determination of the asymptotic effective mass benefits
from a consideration of excited states.
We find that for many of our correlation functions, the behaviour of the
effective mass indeed suggests a late plateau which starts in a region where
the statistical noise is already quite large.
Therefore, estimates based on a single exponential might introduce systematic
uncertainties.

The two-state fits come in two kinds (see
\cref{eq:2state_fit0,eq:2state_fit1}) both of which assume that the
second excited state contribution is negligible in the chosen fit
interval.
Another approximation by Taylor expansion furthermore restricts the
start of the fit interval to values where
$\qty(\expansioncoeff)^2\ll1$.
The isospin-symmetric fits can be done using the exact formula
\cref{eq:2state_fit0_exact}, however, the fit intervals are chosen such that
\cref{eq:2state_fit0} is equally valid. 

As the value for $\Delta M\zth$ in \cref{eq:2state_fit1} is taken from the
isospin-symmetric fit, the fit interval for the isospin-breaking corrections is
chosen such that a zeroth-order fit with the same interval would result in a
value for $\Delta M\zth$ that is compatible with the one used for the
isospin-symmetric fit.

Note that $\Delta M\zth$ does not necessarily correspond to the mass difference
between the ground state and the first excited state as corrections from higher
excited states can in principle be absorbed by this fit value, especially if
the fit interval starts at small values of $t$ where the statistical noise
might be small enough for the effective mass to be sensitive to these
corrections.
Furthermore, the spectrum of excited states is, in general, too dense to be
resolved in a single two-point-function.
$\Delta M\zth$ should therefore not be understood as a physical quantity but
rather as a proxy to parameterize the curvature of the effective mass.

\begin{figure}[htb]
    \centering
    \begin{subfigure}[b]{.49\textwidth}
        \includegraphics[width=\textwidth]{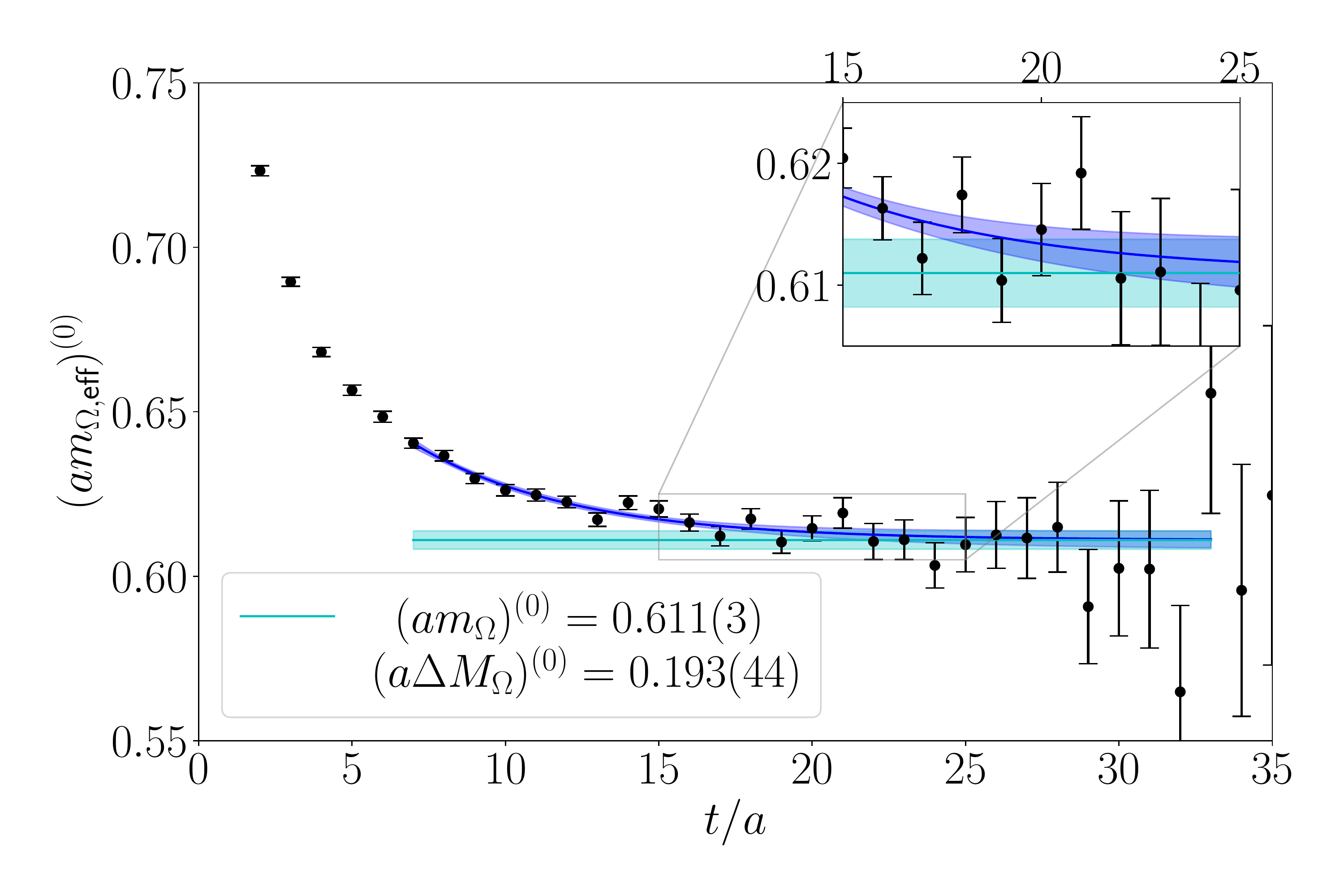}
        \label{fig:omega0}
    \end{subfigure} \\
    \begin{subfigure}[b]{.49\textwidth}
        \includegraphics[width=\textwidth]{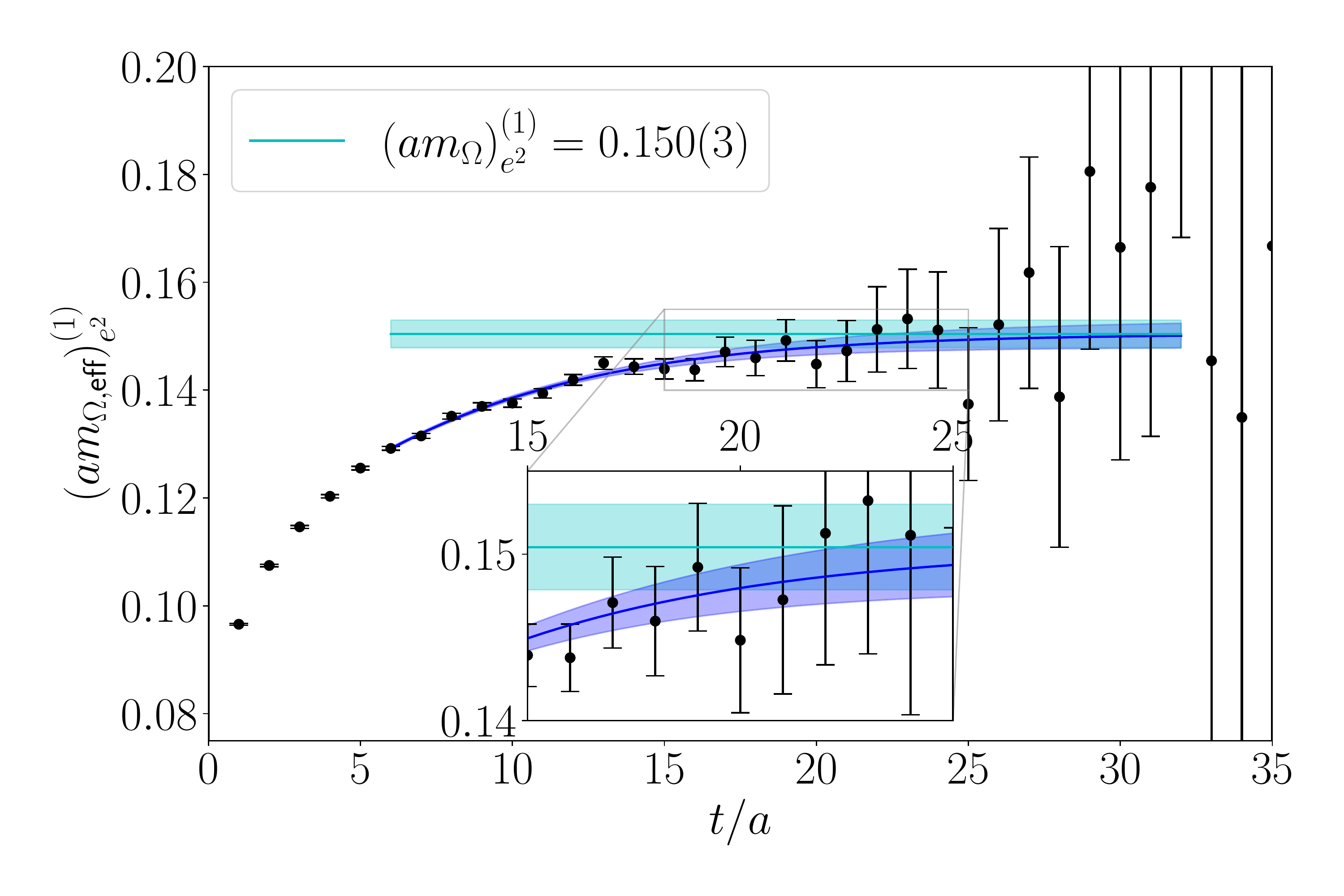}
        \label{fig:omegaqed}
    \end{subfigure}
    \begin{subfigure}[b]{.49\textwidth}
        \includegraphics[width=\textwidth]{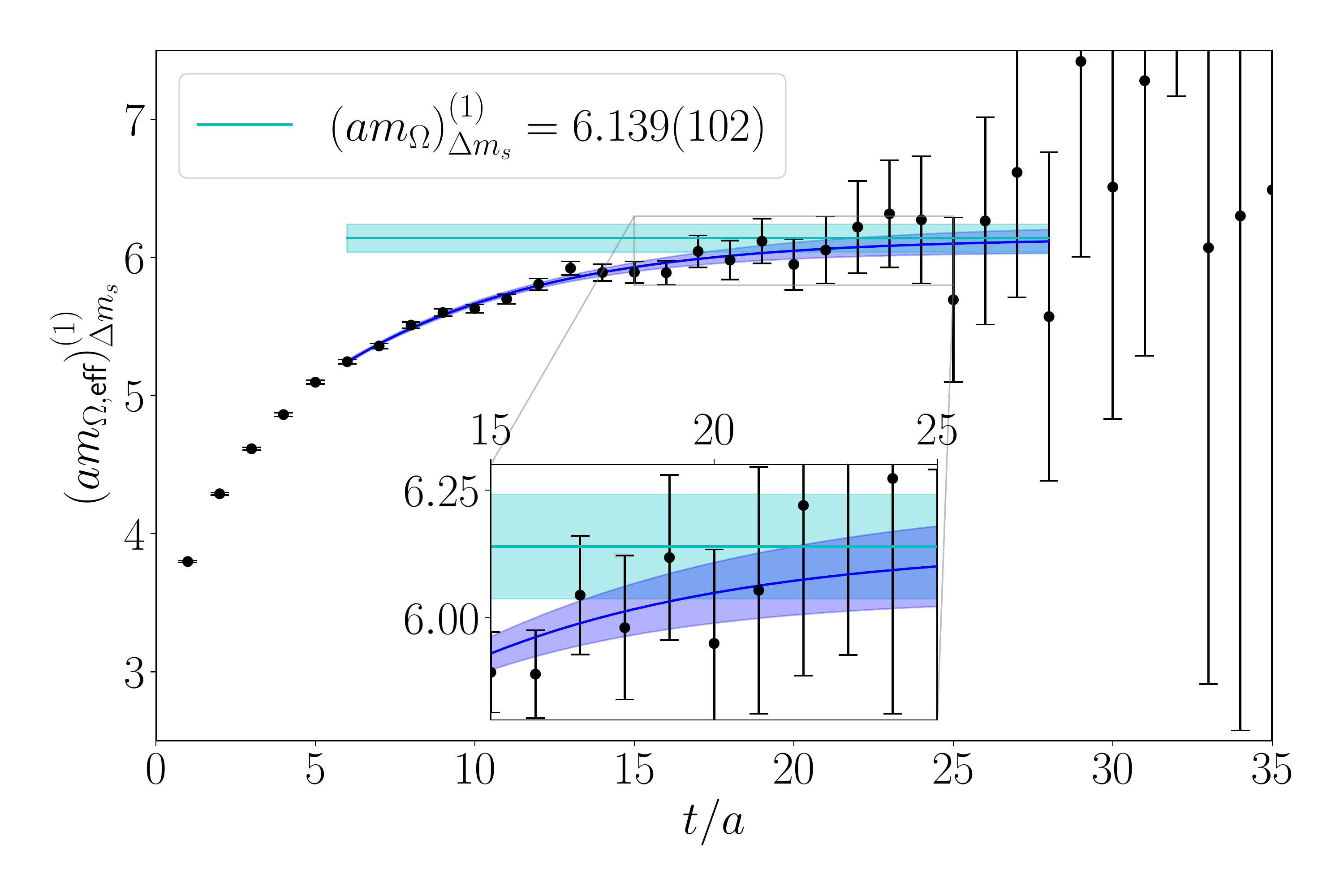}
        \label{fig:omegams}
    \end{subfigure}
    \caption{
        Effective mass plots and fits of the $\Omega$ baryon for the
        isospin-symmetric case (top), the QED corrections (lower left), and the
        $\Delta m_s$ corrections (lower right) on D450.
        The blue bands show the results of two-state fits according to
        \cref{eq:2state_fit0,eq:2state_fit1}.
		Their asymptotic values which correspond to the ground state mass or
		its corrections are shown in cyan.
        The fit interval is indicated by the horizontal extent of the blue and
        cyan lines.
        The effective masses are computed via \cref{eq:meff}.
    }
    \label{fig:omega}
\end{figure}

An example for the fits of all contributions of the $\Omega$-baryon on D450 is
shown in \cref{fig:omega}.
The fit intervals were chosen according to the criteria discussed above. The
smaller insets illustrate the aforementioned problem of a late onset of a
plateau, since the deviation of the fit curve from the asymptotic value is
comparable to the error of the effective mass around $t/a=20$ where the
statistical fluctuations of the first-order correlation functions become
uncontrollably large.
As the $\Omega$ is among the least noisy baryon channels, this problem is often
more severe for other baryons.

By looking at the different members of the baryon octet and decuplet, we
found that the most promising candidates to use for setting the scale in terms
of precision are the $\Omega$ and the $\Xi$.
The values for the isospin-symmetric masses and their corrections we find from
our analysis of N451 and D450 data in lattice units are listed in
\cref{tab:candidates}.

\begin{table}[htb]
    \centering
    \caption{Preliminary results for the most promising candidates for scale
    setting on the CLS ensembles N451 and D450.
    The different columns represent different contributions in the
    isospin-breaking expansion.
    $\delta(am_0)$ denotes the shift of the ground-state mass due to the
    isospin-breaking corrections shown in the other columns.
    To obtain this value we use results for the isospin-breaking parameters
    from another study using the hadronic renormalization scheme described
    in \cite{Risch:2021hty}.
    All uncertainties are purely statistical.
    }
    \label{tab:candidates}
    \begin{tabular}{|c|c|c|c|c|c|l|}
    \hline
    ensemble & baryon & $am_0^{(0)}$ & $(am_0)_{e^2}^{(1)}$ & $(am_0)_{\Delta m_{u/d}}^{(1)}$ & $(am_0)_{\Delta m_s}^{(1)}$ & $\delta(am_0)$ \\
    \hline\hline
\multirow{3}{*}{N451} & $\Omega^-$ & \SI{0.5970\pm0.0023}{} & \SI{0.1430\pm0.0025}{} & - & \SI{5.90\pm0.10}{} & \hphantom{$-$}\SI{0.00020\pm0.00002}{} \\
    \cline{2-7}
 & $\Xi^-$ & \SI{0.4777\pm0.0014}{} & \SI{0.2031\pm0.0018}{} & \SI{3.303\pm0.061}{} & \SI{5.160\pm0.024}{} & \hphantom{$-$}\SI{0.00126\pm0.00003}{} \\
    \cline{2-7}
 & $\Xi^+$ & \SI{0.4777\pm0.0014}{} & \SI{0.4326\pm0.0060}{} & \SI{3.303\pm0.061}{} & \SI{5.160\pm0.024}{} & \SI{-0.00085\pm0.00003}{} \\
    \hline\hline
\multirow{3}{*}{D450} & $\Omega^-$ & \SI{0.6116\pm0.0025}{} & \SI{0.1504\pm0.0026}{} & - & \SI{6.14\pm0.10}{} & \hphantom{$-$}\SI{0.00024\pm0.00007}{} \\
    \cline{2-7}
 & $\Xi^-$ & \SI{0.4876\pm0.0011}{} & \SI{0.2003\pm0.0029}{} & \SI{3.198\pm0.078}{} & \SI{5.077\pm0.032}{} & \hphantom{$-$}\SI{0.00132\pm0.00017}{} \\
    \cline{2-7}
 & $\Xi^+$ & \SI{0.4876\pm0.0011}{} & \SI{0.423\pm0.011}{} & \SI{3.198\pm0.078}{} & \SI{5.077\pm0.032}{} & \SI{-0.00076\pm0.00060}{} \\
    \hline
\end{tabular}
\end{table}

Our estimates for the mass shifts are based on results from a previous
calculation \cite{Risch:2018ozp,Risch:2021hty}, where the isospin breaking
parameters $e^2=4\pi\alpha_{\text{em}}$, $\Delta m_d$, $\Delta m_u$, and
$\Delta m_s$ were determined by identifying the following quantities in QCD+QED
with their experimentally measured values:
\newcommand{\equalize}[1]{\qty(#1)^{\text{QCD+QED}}=&\qty(#1)^{\text{experiment}}}
\newcommand{\msq}[1]{m_{#1}^2}
\begin{align}
\begin{aligned}\label{eq:expansion_matching}
    \equalize{\msq{\pi^0}}, \\
    \equalize{\msq{K^+}+\msq{K^0}-\msq{\pi^+}}, \\
    \equalize{\msq{K^+}-\msq{K^0}-\msq{\pi^+}+\msq{\pi^0}}, \\
    \equalize{\alpha_{\text{em}}}.
\end{aligned}
\end{align}

These conditions must be evaluated for each ensemble individually, and in order
to perform this matching, a lattice scale has to be used in order to calculate
the masses in physical units from the lattice results.
As we have no results for the lattice scale including isospin-breaking
corrections thus far, the masses computed on the lattice were scaled with a
value from purely isospin-symmetric results.
The results from this matching which are used to compute $\delta(am_0)$ in
\cref{tab:candidates} are
\begin{center}
\begin{tabular}{c||c|c|c}
    Ensemble & $a\Delta m_d$ & $a\Delta m_u$ & $a\Delta m_s$ \\
    \hline
    \hline
    N451 & \SI{-0.0018374\pm0.0000071}{} & \SI{-0.0088474\pm0.0000075}{} & \SI{-0.0021890\pm0.0000018}{} \\
    \hline
    D450 & \SI{-0.0018160\pm0.0000073}{} & \SI{-0.0088457\pm0.0000082}{} & \SI{-0.0022123\pm0.0000061}{} \\
\end{tabular}.
\end{center}

For the baryons listed in \cref{tab:candidates} we find that on the given
ensembles we reach a precision of better than \SI{0.5}{\percent}, with the size
of isospin-breaking corrections being of the same order as the zeroth-order
uncertainty or even smaller.
However, these uncertainty estimates are purely statistical and systematics
still have to be investigated for a realistic estimate of the uncertainties.
Since we do not take corrections in the sea-quark sector into account, for
example, the listed isospin-breaking contributions are incomplete which has to
be considered as a systematic uncertainty.
Furthermore, isospin-breaking corrections of the lattice scale have been
neglected in the matching in \cref{eq:expansion_matching}, and finite-volume
effects were not considered in any of our results presented here.

For the $\Omega$ baryon we find only very small mass shifts arising from
isospin-breaking effects.
This is not surprising and is one of the reasons this baryon is generally a
popular candidate for scale-setting
\cite{RBC-UKQCD:2008mhs,Borsanyi:2020mff,Miller:2020evg}.


\section{Conclusion and Outlook}\label{sec:conclusion}

We have presented our method for the determination of baryon octet and decuplet
masses and their corrections due to isospin-breaking effects from correlation
functions computed as described in \cite{Segner:2021yqo}.
We found that small statistical uncertainties on these quantities can be
achieved for the $\Omega$ and $\Xi^\pm$ baryons which we therefore consider as
prime candidates for setting the scale of CLS $N_f=2+1$ ensembles.
As we have only processed two ensembles so far, we are currently extending our
analysis to additional ensembles, in order to thoroughly investigate the
suitability of our baryonic mass estimates for scale setting.

\acknowledgments

The authors gratefully acknowledge the Gauss Centre for Supercomputing e.V.
(www.gauss-centre.eu) for funding this project by providing computing time
through the John von Neumann Institute for Computing (NIC) on the GCS
Supercomputer JUWELS at Jülich Supercomputing Centre (JSC).

The work of ADH is supported by: (i) The U.S. DOE, Office of Science, Office of
Nuclear Physics through Contract No. DE-SC0012704 (S.M.); (ii) The U.S. DOE,
Office of Science, Office of Nuclear Physics and Office of Advanced Scientific
Computing Research within the framework of Scientific Discovery through
Advanced Computing (SciDAC) award Computing the Properties of Matter with
Leadership Computing Resources.

\providecommand{\href}[2]{#2}\begingroup\raggedright\endgroup

\end{document}